\documentclass[prl,showpacs,twocolumn]{revtex4}

\usepackage{graphicx}

\begin{document}
\bibliographystyle{prsty}

\title{Surface Plasmon Polariton microscope with Parabolic Reflectors}

\author{Aurelien Drezet}
\author{Daniel Koller}
\author{Andreas Hohenau}
\author{Alfred Leitner}
\author{Franz R. Aussenegg}
\author{Joachim R. Krenn}

\affiliation{Institute of Physics and Erwin Schr\"{o}dinger
Institute for Nanoscale Research, Karl--Franzens--University,
Universit\"{a}tsplatz 5, 8010 Graz, Austria}

\date{\today}

\begin{abstract}
We report the realization of a two--dimensional optical microscope
for surface plasmons polaritons (SPPs) based on parabolic Bragg
mirrors. These mirrors are built from lithographically fabricated
gold nanostructures on gold thin films. We show by direct imaging
by leakage radiation microscopy that the magnification power of
the SPP microscope follows basic predictions of geometrical
optics. Spatial resolution down to the value set by the
diffraction limit is demonstrated.
\end{abstract}

\maketitle

Surface plasmon polaritons (SPPs) are propagative electromagnetic
waves bound to the interface between a metal and a dielectric
\cite{Raether}. As such, SPPs offer several possibilities for
two--dimensional optical functionalities at micro- and nanometer
scales \cite{Barnes:2003}. Various quasi two--dimensional SPP
elements were demonstrated recently such as waveguides
\cite{Weeber-Lacroute-Dereux:2003,Nikolajsen:2003,Harry2005},
lenses \cite{Hohenau,Liu}, reflectors \cite{Harry,Yin} and
interferometers \cite{Harry,Drezet,sergey}. In particular, the
realization of an elliptical interferometer for SPPs based on
confocal Bragg reflectors built on a flat gold film was reported
\cite{Drezet}. In this interferometer SPPs launched from the first
focal point $F_{1}$ are efficiently focussed in the second focal
point $F_2$ after propagating inside an elliptical cavity. Clearly
such a geometry can as well be used for in--plane microscopic
imaging \cite{limitation}. In fact, SPP microscopy has been a
topic of increased interest recently, including some controversy
about the achievable spatial resolution \cite{smolya,comment}. For
geometrical reasons, however, an elliptical geometry does not
allow to obtain an image magnification much different from unity.
To overcome this limitation we report here the realization of an
SPP in--plane microscope with parabolic mirrors. We realize
two--dimensional SPP imaging with variable magnification and show
that a spatial resolution of about half the SPP wavelength is
attainable in agreement with the Abbe-Rayleigh criterion
\cite{Teich,Andrey2}.

All structures discussed in the following are built from gold
nanoprotrusions (150 nm diameter, 70 nm height) fabricated by
standard electron--beam lithography \cite{M3:1997} on a 70 nm
thick gold thin film on glass substrate. To achieve SPP
reflectivity close to unity a Bragg mirror geometry has been
chosen for the reflectors constituted by 11 parabola sharing the
same focal point (and having thus slightly different focal
lengths) and the same symmetry axis. The individual parabola are
built from closely spaced protrusions (center--to--center distance
 $\sim 250$ nm). Constructive interference occurs between the various
reflected and refocussed SPP waves if the closest separation
between two neighboring parabola (measured along the symmetry
axis) equals $\lambda_{\textrm{SPP}}/2$, where
$\lambda_{\textrm{SPP}}$ is the SPP wavelength which is here fixed
to 785 nm, corresponding to a laser excitation wavelength of 800
nm.

As a first example we consider a confocal geometry composed of two
parabolic reflectors having the same focal length and common
symmetry axis $F_1 F_2$ (see Fig.~1). The object to be imaged is
located close to $F_{1}$. From the geometrical properties of
parabola mirrors one can see that the light path $F_1M +MH$ (where
$M$ is the effective SPP reflection point on the parabola and $H$
is the horizontal projection of $M$ onto a line perpendicular to
the parabola axis, see Fig.~1A) is independent of the choice of
the position of $M$. Clearly this means that $F_1$ is projected
into a SPP beam propagating along $F_1 F_2$ and finally that a SPP
launched in $F_{1}$ is refocussed in $F_2$.

In Fig.~1B the object in $F_1$ is a single gold nanoprotrusion.
SPPs are launched on the metal film by focussing on this
nanoprotrusion a laser beam impinging perpendicularly to the
surface with a polarization direction oriented along the vertical
direction. The imaging of SPP propagation is accomplished by
leakage radiation microscopy (LRM) which relies on the conversion
of SPP waves into propagating light through the glass substrate
\cite{Hecht,Bouhelier,Andrey}. It is clearly visible in the image
that efficient SPP refocussing is obtained in $F_2$, with an
intensity pattern around the focus that corresponds to that in a
plane containing the optical axis in conventional microscopy. We
note that a large part of the image is saturated due to the
limited dynamics of the camera used to capture the LRM images.
Figs.~1C and D show LRM images for the case of a protrusion pair
as the object in the area of $F_1$. Here, the pair axis is aligned
along the symmetry axis of the mirrors, i.e., along the horizontal
direction and the protrusion distances are $d=1\mu$m (C) and
$d=2\mu$m (D), respectively. The insets in Figs.~1B-D show
intensity cross--cuts along the symmetry axis around $F_2$ as
marked by the triangles in the images. In both cases the object
structure is clearly imaged in $F_2$. The measured separations of
the image spots correspond to those of the protrusion pairs,
directly demonstrating unity longitudinal magnification for the
case of identical focal distances of the two parabola.

We now turn to image magnification by using two different parabolic
reflectors with focal lengths $f_1=5\mu$m and $f_2=15\mu$m (see
Fig.~2A). Fig.~2B shows a two--dimensional dipolar simulation
\cite{Theseharry} of the expected intensity distribution in the
parabolic cavity when SPPs are launched from a single nanoprotrusion
located at $F_1$.

This result reproduces fairly the experimental LRM image
corresponding to this configuration, shown in Fig.~2C. As
previously reported, in LRM certain SPP components with specific
in--plane momentum can be Fourier filtered in the back focal plane
of the imaging microscope objective \cite{Drezet2}. Thereby these
components and according spurious interference can be eliminated
from the direct space images , exposing otherwise obstructed
features to direct analysis. Fig.~2D zooms a detail of Fig.~2C
after filtering all SPP components propagating from right to left.
As now only SPPs reflected by the second (left) mirror, i.e.,
propagating from left to right contribute to the image we recover
a clear image of the focal region.

For demonstrating microscopic magnification we now analyze the
transverse magnification of our SPP in--plane microscope
\cite{limitation} which is simply given by $f_2/f_1=3$. We
therefore consider again nanoprotrusion pairs located near $F_1$.
While the position of one protrusion is in $F_1$, the second
protrusion is separated by a distance along the direction
perpendicular to the microscope symmetry axis of $d'=2$, $1$ and
$0.4$ $\mu$m (see Fig.~3). The according LRM images are shown in
Figs. 3A, C and E which again display Fourier filtered images,
suppressing all SPP components propagating from right to left.

The images of the two protrusion are clearly visible and resolved,
even for the smallest distance
$d'=0.4\mu$m$\simeq\lambda_{\textrm{SPP}}/2$.
Using the filtered LRM images we obtain transverse cross--cuts of
the SPP intensity along the line joining the centers of the two
image points, see panels B, D and F of Fig.~4. To guide the eye we
include Gaussian fits to the data. Clearly the image of the
off--axis protrusion is broader than that of the on--axis
protrusion which is due to abberation. Most importantly, however,
the observed distances between the image spots are a factor of $3$
larger than the distances between the nanoprotrusions, thus
directly evidencing the expected factor for the transverse
magnification as expected from geometrical optics.

To conclude, we have experimentally demonstrated an in--plane SPP
microscope relying on parabolic Bragg mirrors with a spatial
resolution of about $\lambda_{\textrm{SPP}}/2$. In the context of
plasmonics this microscopic scheme could complement existing
elements as waveguides, mirrors, etc., for the benefit of
applications in SPP based imaging, optical sub--wavelength
addressing or sensing.

\newpage
Figure: 1//

Confocal parabolic SPP microscope. (A) Scanning electron
microscope image. $F_1$ and $F_2$ are the focal points, the focal
distances are $f_1=f_2= 15 \mu$m and the lines show exemplary SPP
paths from $F_1$ to $F_2$. The inset displays a magnified image of
the nanoprotrusions constituting the Bragg mirror structure. The
minimum separation between two adjacent mirrors is
$a=\lambda_{\textrm{SPP}}/2=390 nm$. (B) LRM image with a single
nanoprotrusion at $F_1$. The laser polarization is indicated by
the double arrow. (C) Same as (B) but with a protrusion pair
separated by a horizontal distance $d=1 \mu$m. (D) Same as (B) but
with a protrusion pair separated by a horizontal distance $d=2
\mu$m. The insets in (B)--(D) show cross--cuts of the SPP
intensity along the symmetry axis near $F_2$, as indicated by the
triangles in the images. The cross--cut scale bar is 2 $\mu$m, the
vertical lines are separated by d=1 $\mu$m (C) and 2 $\mu$m (D).

Figure 2: //

Magnifying parabolic SPP microscope. (A) Scanning electron
microscope image. (B) Two--dimensional dipolar simulation of SPPs
propagating inside the parabolic mirrors after launching from
$F_1$. The laser polarization direction is indicated by the double
arrow. (C) LRM image corresponding to (B). (D) Magnified LRM image
corresponding to the dashed white rectangle in (C) after Fourier
filtering of the SPP components propagating from left to right.

Figure 3: //

Microscopic imaging of a nanoprotrusion pair located near $F_1$
using the geometry shown in Fig.~2. A, C and E show the Fourier
filtered LRM images for pairs (SEM images shown in the  insets)
separated by $d=2$, $1$ and $0.4$ $\mu$m, respectively. The images
cover areas corresponding to the same dashed rectangles as in
Fig.~2C.B, D, and F show LRM intensity cross--cuts taken along the
dashed lines of images A, C and E, respectively. The curves are
Gaussian fits to the data.

\newpage
Figure 1://

\begin{figure}[h]
\includegraphics[width=8cm]{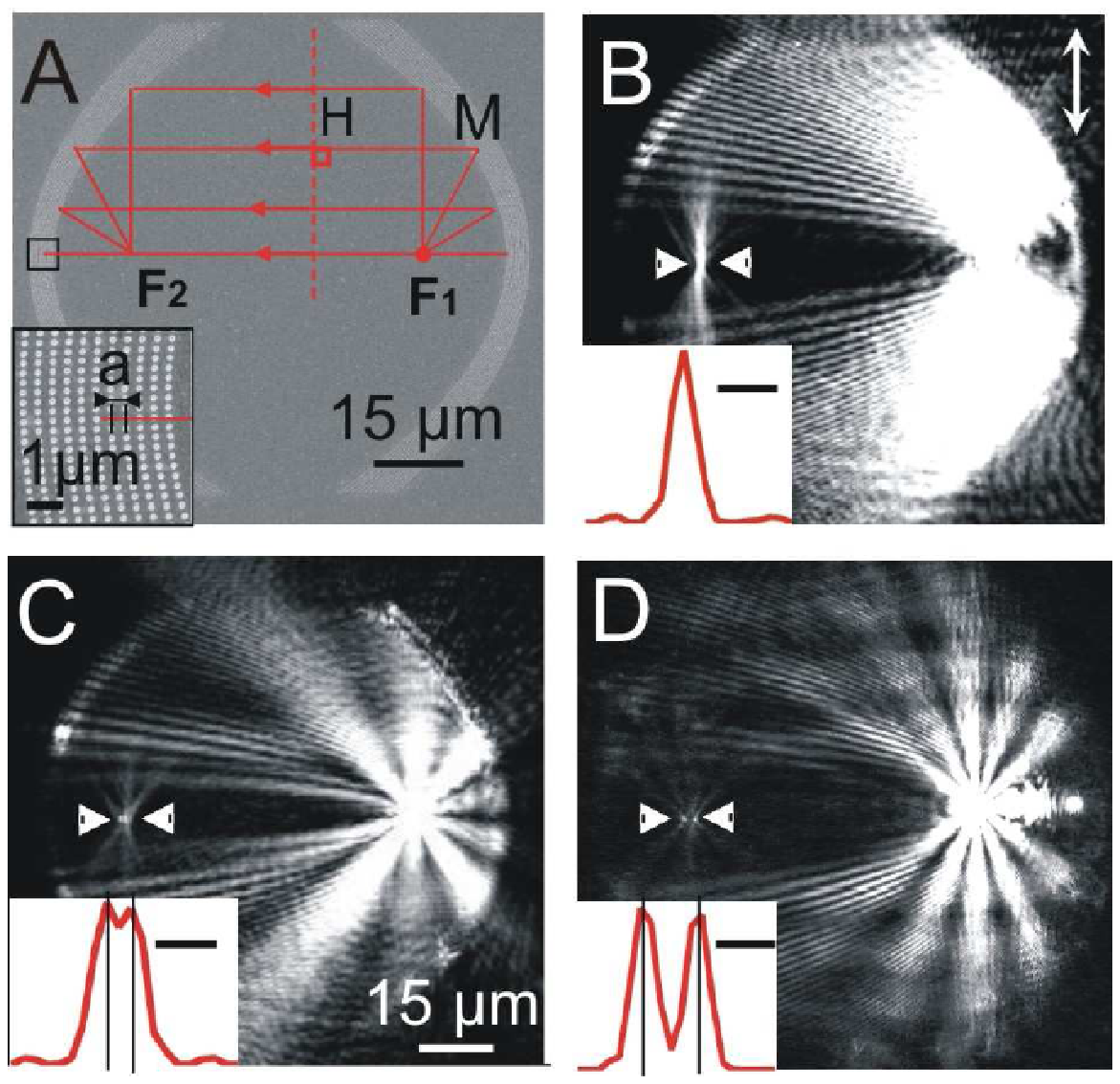}
\caption{}
\end{figure}

\newpage
Figure 2://

\begin{figure}[h]
\includegraphics[width=8cm]{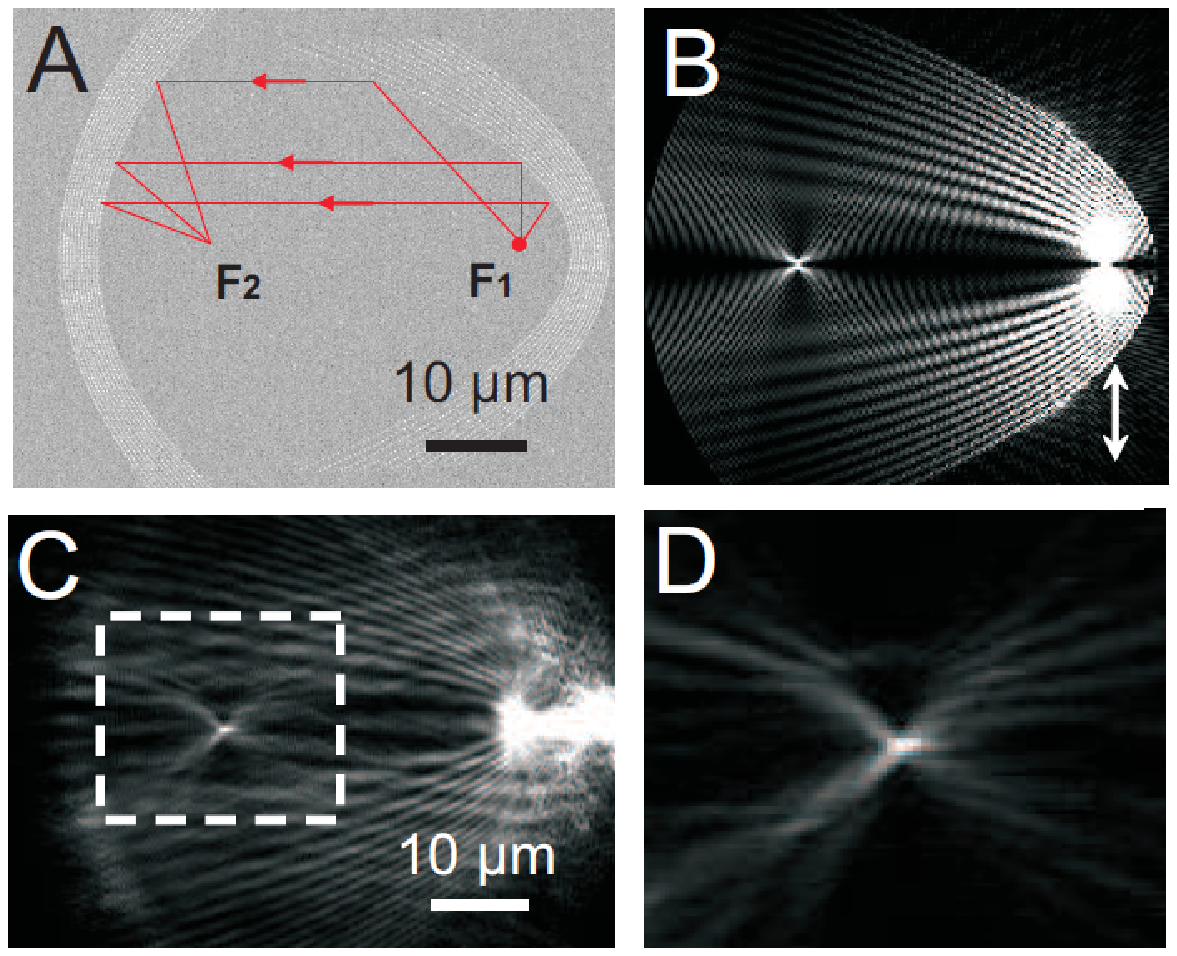}
\caption{}
\end{figure}

\newpage
Figure 3://

\begin{figure}[h]
\includegraphics[width=8cm]{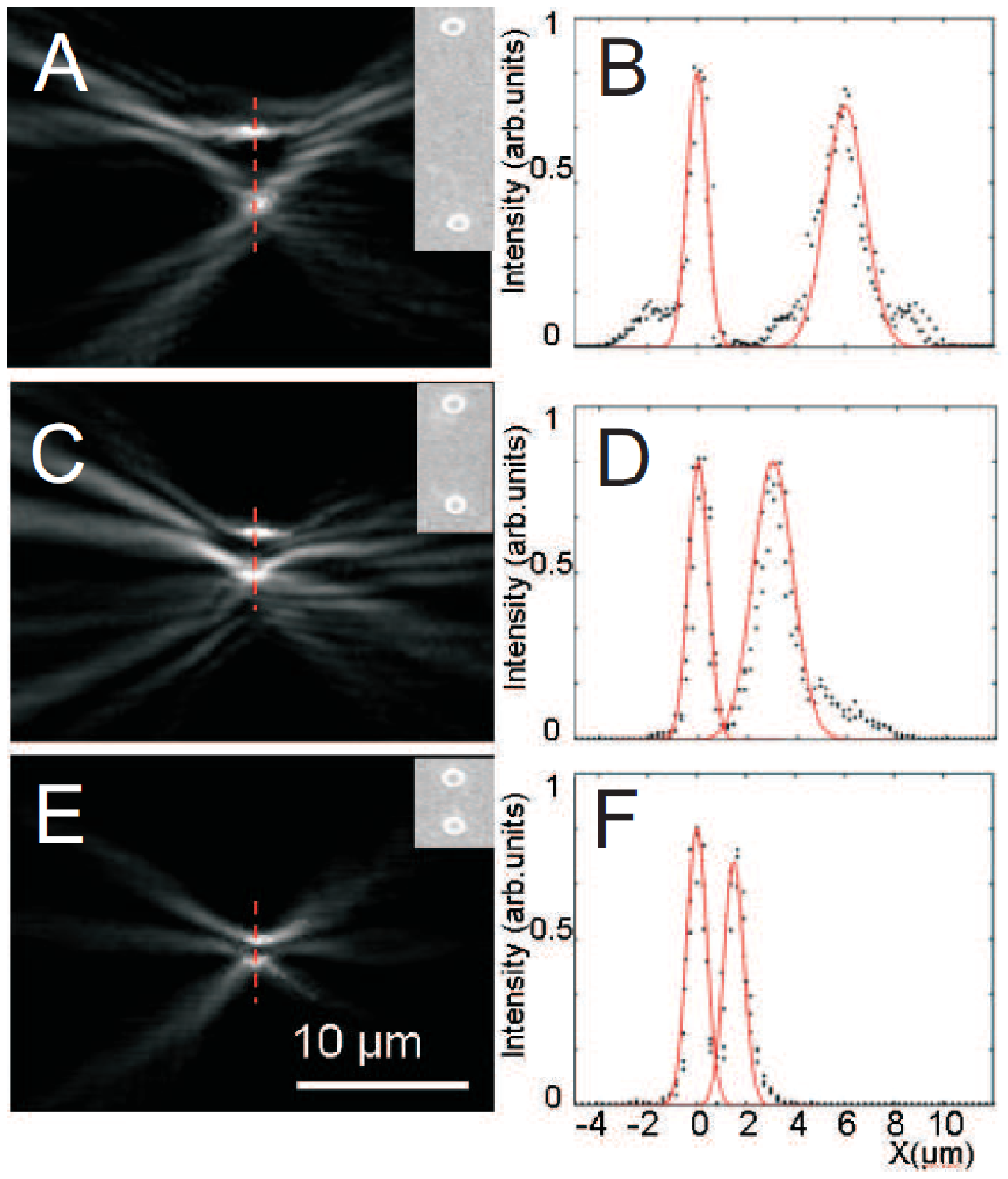}
\caption{}
\end{figure}
\end{document}